\documentclass[reprint, amsmath, amsfonts]{revtex4-1}
\pdfoutput=1
\def\minus{%
  \setbox0=\hbox{-}%
  \vcenter{% 
    \hrule width\wd0 height \the\fontdimen8\textfont3%
  }%
}

\usepackage{graphicx}
\usepackage{amssymb}
\usepackage{wasysym}
\usepackage{stmaryrd}
\usepackage{dcolumn}

\usepackage{lineno,hyperref}
\usepackage{amsmath,bm}

\newcommand{\etat}{\eta_\text{\tiny T}}

\newcommand{\Rm}{R_\text{\tiny min}}
\newcommand{\kr}{k_\text{\tiny R}}
\newcommand{\al}{\alpha_\text{\tiny r}}
\newcommand{\alv}{\alpha_\text{\tiny v}}

\begin{document}

\title{Perfect absorption of water waves by linear or nonlinear critical coupling}

\author{E. Monsalve}
\email{eduardo.monsalve@espci.fr}
\affiliation{Laboratoire de Physique et M\'{e}canique des Milieux H\'{e}t\'{e}rog\`{e}nes, CNRS, ESPCI-Paris, PSL Research Univ., Sorbonne Univ., Univ. Paris Diderot, 10 rue Vauquelin, 75005 Paris, France}
\author{A. Maurel}
\affiliation{Institut Langevin, CNRS, ESPCI-Paris, 1 rue Jussieu, 75005 Paris, France}
\author{P. Petitjeans}
\affiliation{Laboratoire de Physique et M\'{e}canique des Milieux H\'{e}t\'{e}rog\`{e}nes, CNRS, ESPCI-Paris, PSL Research Univ., Sorbonne Univ., Univ. Paris Diderot, 10 rue Vauquelin, 75005 Paris, France}
\author{V. Pagneux}
\affiliation{Laboratoire d\'{}Acoustique de l\'{}Universit\'{e} du Maine, CNRS, Avenue Olivier Messiaen, 72085 Le Mans CEDEX 9, France}

\date{\today}

\begin{abstract}
We report on experiments  of  perfect absorption for  surface gravity   waves impinging a wall structured by a subwavelength resonator.
 By tuning the geometry of the resonator, a balance is achieved between the radiation damping and the intrinsic viscous damping, resulting
 in perfect absorption by critical coupling. 
 Besides, it is shown that the resistance of the resonator, hence the intrinsic damping, can be controlled
 by the wave amplitude, which provides a way for 
perfect absorption tuned by nonlinear mechanisms. 
The perfect absorber that we propose, without moving parts or added material, is simple, robust and it presents
a deeply subwavelength ratio wavelength/size $\simeq 18$.
 
 \end{abstract}

\maketitle

Waves are generically absorbed when they interact with a resonator close to the resonance frequency.
To increase the absorption for scattering problems, a useful idea is that of critical coupling which aims at a balance between 
damping  different in kind:  radiation damping and intrinsic damping.
Indeed, a resonator coupled to an infinite domain undergoes radiation damping \cite{carrier1971response,mei_theory_2005} 
even in the absence of intrinsic losses that would be present for a closed isolated resonator (e.g. viscous losses).
When the balance is realized,  waves generated by radiation damping and by  intrinsic damping interfere destructively 
which results in perfect absorption: the incident wave does not generate any scattered wave. 
This concept has been successfully applied in the field of absorption in  electromagnetism \cite{cai2000,yariv2002,landy_perfect_2008,watts2012metamaterial,chong_coherent_2010,luk_directional_2014,fan2014} and in acoustics 
%\cite{ma_acoustic_2014,sheng2016,sheng2017,romero-garcia_perfect_2016,long2017,lee2018} for instance.
%romero-garcia_use_2016

 In the context of water waves, 
 devices able to absorb or to extract  the energy from sea waves  have been foreseen for a long time \cite{mei_theory_2005}. 
  Primary wave-energy devices consisted in oscillating floating 
    \cite{evans_theory_1976,mei1976power}
  or 
  submerged bodies  
  \cite{simon_wave-energy_1981,crowley2013submerged},
    for a review see \cite{evans_wave_2012,falnes_review_2007}. 
    The need of the reduction of reflection of waves in harbours and basins has also attracted interests \cite{twu,theocharis}.
    More recently, new devices have been sought considering the opportunity to combine them with existing breakwaters on the coast
    (see {\em e.g.} \cite{martins-rivas_wave_2009}). 
 In most of the cases,  the mechanism of energy absorption relies on  
 resonances.

In this work, we show experimentally that  the use of the critical coupling concept is an easy way to achieve total absorption. 
We choose a simple resonator with a tunable geometry allowing us to cover a large range of  radiation damping.
In order to obtain critical coupling, the geometry of the resonator is tuned
until the right balance between the radiation damping and the inherent viscous losses is achieved.
Besides, we show that we can take advantage of  the wave nonlinearities, which is the rule rather than the exception for sea waves, to  tune  the absorption toward the critical coupling.  
\begin{figure}[t]
\centering 
	\includegraphics[width=.8\columnwidth]{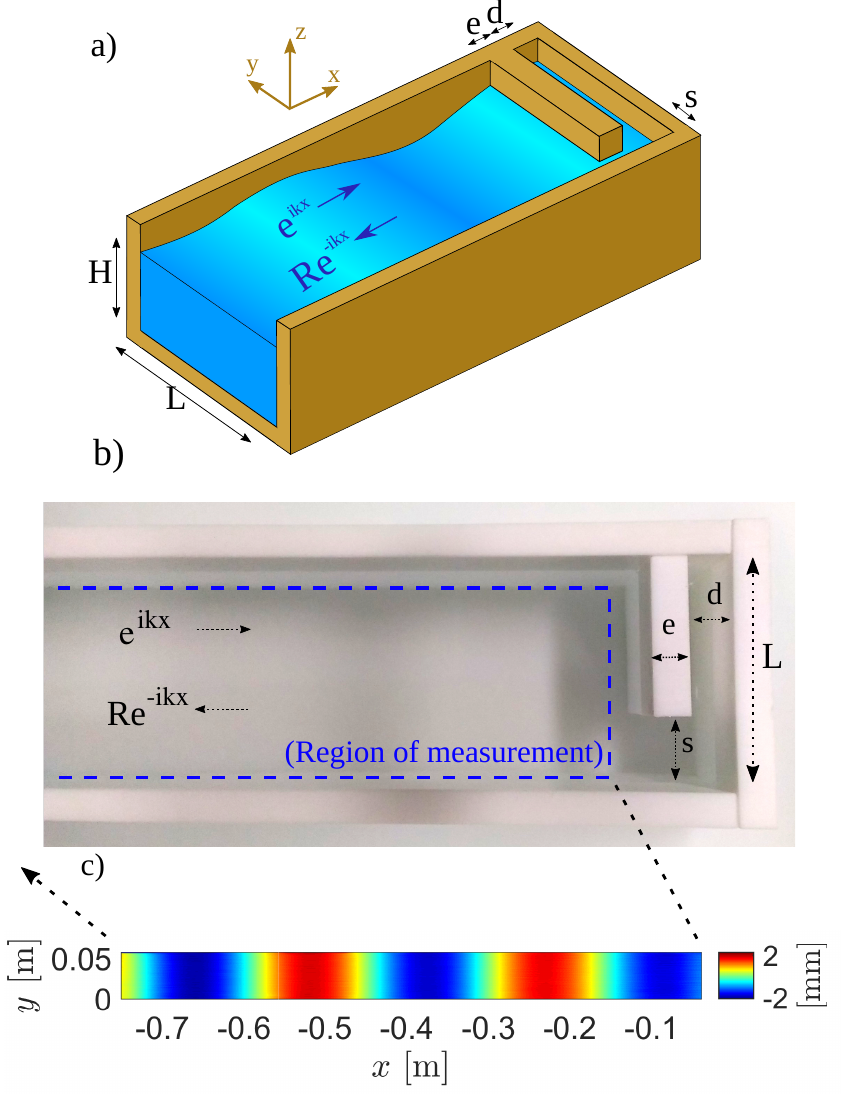}%
 \caption{ 
 Experimental set-up of the subwavelength resonator absorber for water waves in a channel. 
 a) 3-dimensional scheme. 
 b) Top view of the end part of the channel with the resonator.
  c) Typical surface elevation wave field  measured experimentally in the channel with the FTP technique.}
	\label{fig:Fig1_two_columns}
\end{figure}

The resonator that we consider is composed of a small open cavity of depth $d$ at the extremity of a channel of width $L$ (Fig. \ref{fig:Fig1_two_columns}). 
A gate of width $e$ delimits  the open cavity connected to the channel through a thin guide corresponding to the resonator neck with
an opening length $s$. 
By varying the opening $s$, the radiation damping of the resonator will change a lot:
for $s=L$ it is totally open to leakage and for $s \rightarrow 0$ it will go to a closed resonator without leakage.
Since the bathymetry is flat, with a finite depth $H$, the surface elevation $\eta$ satisfies the Helmholtz equation 
\begin{equation}
\Delta \eta+k^2\eta=0,
\end{equation}
in which for a given frequency $\omega$ (harmonic regime $e^{-i\omega t}$), the wavenumber is given by the dispersion relation of water waves
\begin{equation}
\label{disp}
\omega^2=g\,k \tanh kH,
\end{equation}
with $g$ the gravity and $H$ the water depth. 
In the following, we will be working in the low frequency regime, with $kL<\pi$, where only the planar mode can propagate in the channel.
An incident wave  impinges on the resonator which is at $x=0$, such that
 the total wave in the channel can be written in the form
 \begin{equation}
 \eta(x)=a( e^{ikx}+Re^{-ikx}), \quad x<0,
 \label{eqeta}
 \end{equation}
 with $a$ the complex amplitude of the incident wave at $x=0$ and
where $R$ is the reflection coefficient. 

\begin{figure}[h]
\centering 
	\includegraphics[width=.8\columnwidth]{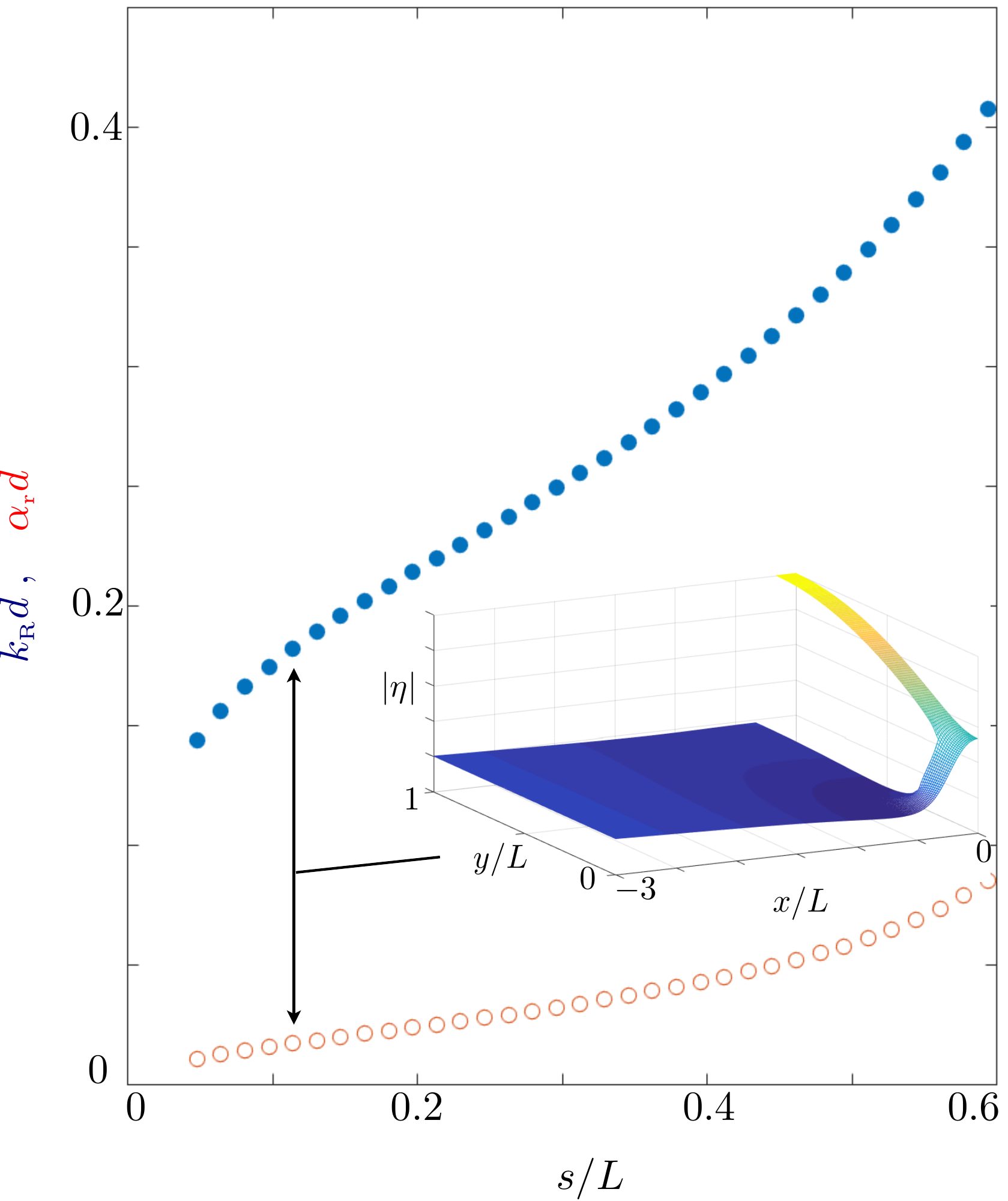}%
 \caption{ Numerically computed complex resonances:
 Dimensionless real part $\kr d $ (filled blue) and imaginary part $\al d$ (empty red) of the complex resonance frequency as a function of the opening of the
 resonator $s/L$. The inset shows the shape of the absolute value of the elevation $|\eta|$ at resonance frequency for $s/L=0.11$.}
	\label{figreso}
\end{figure}

We first consider the lossless case; then, close to the resonance frequency, $R$ can be expressed as
\begin{equation}
R=\frac{k-\kr - i \al }{k-\kr  +i \al },
\label{eqR0}
\end{equation}
where $\kr  -i \al$ corresponds to the complex resonance frequency of the resonator \cite{pagneux1}, 
where $\al$ encapsulates  the radiative damping.
This complex resonance frequency has been computed numerically
 (with the pdetool Finite Element Method toolbox of Matlab) as a function of the resonator opening $s/L$; the results 
are shown in Fig. \ref{figreso}. 
We can observe that $\al$ spans a broad range of values and it is thus confirmed that
the chosen simple resonator offers a wide range of radiation damping. It is an important 
aspect because it means that it is a factor that we will be able to tune easily.

Then, a good approximation to take into account the intrinsic loss in the cavity can be obtained by simply shifting the numerator and the denominator
by a  positive viscous damping factor $\alv$ (as if  $k\to k+i\alv$) %\cite{romero-garcia_use_2016}:
\begin{equation}
R=\frac{k-\kr - i(\al -\alv) }{k-\kr  +i (\al +\alv)},
\label{eqR}
\end{equation}
We shall see that the  above expressions \eqref{disp} and  \eqref{eqR} 
 contain all the necessary to describe the critical coupling in our experiments.
In every case, the point will be to try to obtain $\al =\alv$ to get $R(k=\kr)=0$, the total absorption at
the resonance frequency.

\vspace{.3cm}

\noindent {\em Experimental set-up and measuring technique} -- 
The  water depth is set to $H=5$ cm; the channel is  $L=6.2$ cm large and  1 m long. The dimensions of the resonator are  $d=e=$ 1 cm and variable opening $s$ from 0.3 to 3 cm.  
 Waves are generated by  a piston-type wavemaker driven by a linear motor in the range $f\in (1,3.5)$ Hz with   step $0.01$ Hz; their amplitudes $a$ are controlled precisely from 0.2 mm to  5 mm with step 0.1 mm.
 In this frequency range, only the plane wave is propagating ($k\in (10,45)$ m$^{-1}$, whence $kL<\pi$). 
The  fields of surface elevation  $\eta(x,y,t)$ are measured using the optical Fourier Transform Profilometry (FTP) \cite{cobelli_global_2009,maurel2009experimental,przadka2012fourier}. Sinusoidal fringes are projected by a digital projector over an area of $72\times5.7$ cm$^2$. The images of the free surface are collected by a camera, allowing for a spatial resolution  given by the size of the projected pixel of 0.7 mm in both directions and a temporal resolution of $50$ fps given by the acquisition  frequency of the camera. The record duration of  $16$ s covers at least 20 periods for the lowest forcing frequency.
From these space-time resolved measurements of the surface elevation $\etat$, 
we extract the linear mode from a temporal Fourier decomposition 
\begin{equation}
\eta(x,y)=\frac{1}{t_f}\int_0^{t_f} \etat(x,y,t) e^{i\omega t}dt,
\label{eq:Fourier}
\end{equation}
where $\omega$ is the forcing frequency and $t_f=2n\pi/\omega$ with $n$ integer.
This allows us to quantify the nonlinearity of the waves and to calculate  the reflection coefficient $R$, far enough from the resonator (about $4$ cm in practice)  by using a fit of equation \eqref{eqeta}.
Eventually, the signal-to-noise ratio is reduced by averaging  $\eta(x,y)$ 
over the transverse direction $y$ afterwards the fit is performed to get  $(a,R,k)$. Note that the typical measured  field reported in Fig. \ref{fig:Fig1_two_columns} shows that $\eta$ is indeed independent of $y$.

\vspace{.3cm}

\begin{figure}[h!] 
	\centering
	\includegraphics[width=.8\columnwidth]{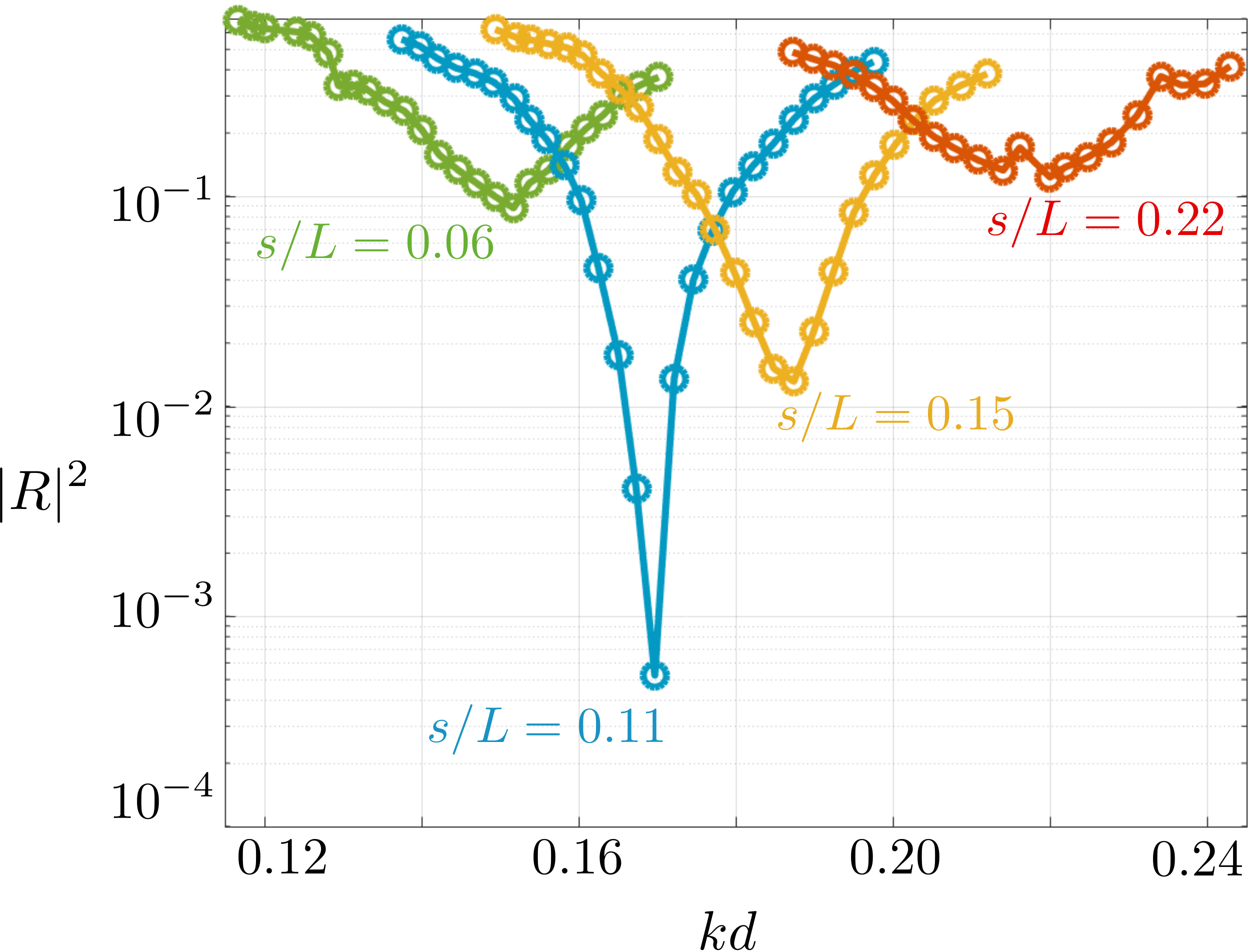}%
\caption{Experimentally measured 
reflection $|R|^2$ against $kd$ in the linear regime for  different radiative damping  dictated by the neck opening $s/L$ of the resonator.
The corresponding resonance frequencies $k_R d$ computed numerically for the lossless case and displayed in 
Fig. \ref{figreso} are $k_R d = 0.15; 0.18; 0.20; 0.22$ for $s/L= 0.6; 0.11; 0.15; 0.22$. }
	\label{fig:graph_reflection_linear}
\end{figure}

\noindent {\em Critical coupling tuned by the geometry} -- 
To begin with, we consider the linear regime for water waves and play with the opening $s$ of the resonator to modify the critical coupling, trying to
push $\al$ towards $\alv$ in \eqref{eqR}). In our experiments,  linear regime  corresponds to  wave amplitudes $a\sim 0.5$ mm,  where we found the nonlinearities to be weak and where our measurement technique 
is accurate. 
The Fig. \ref{fig:graph_reflection_linear} 
reports  the measured reflectivity  $|R|^2$ for 
4 values of  $s/L$ from 0.06 to 0.22. In each case, the reflectivity is smaller than unity because of the viscous losses and it has  a minimum at the  resonance frequency, in rough agreement with \eqref{eqR}. As can be seen, this resonance frequency  is controlled by   the neck opening  $s$, resulting in the shift to higher frequencies when increasing $s$, in agreement with the numerical results in Fig. \ref{figreso}.

\begin{figure}[h!]
	\centering
	\includegraphics[width=.8\columnwidth]{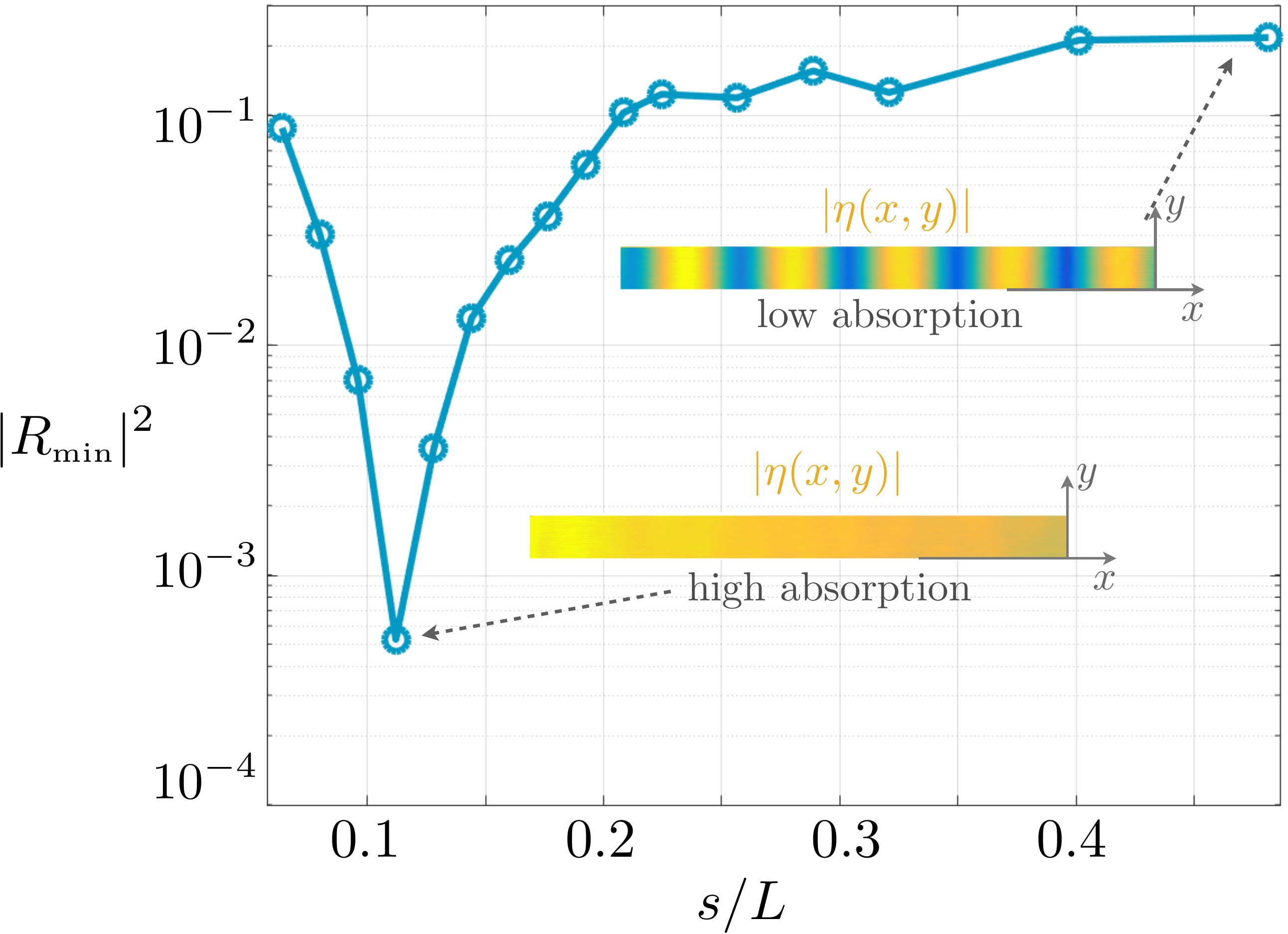}
	 \caption{Critical coupling in the linear regime -- Minimum reflection $|\Rm|^2$ against normalized opening $s/L$, with  critical coupling realized at $s/L=0.11$. The insets show the measured pattern of $|\eta(x,y)|$ in a case of low absorption (revealing interferences between the incident and reflected waves) and in a case of high absorption (with almost constant modulus demonstrating no reflection).}
	\label{fig:graph_R_min_vs_s_linear}
\end{figure}

More interestingly, the minimum in the reflectivity $|\Rm|^2$  is   impacted by the value of the opening $s$: this is what we use
to achieve  the critical coupling with zero reflectivity. 
To reach the total absorption, we perform a series of experiments varying  $s$ with a spacing of $1$ mm and collect the $\Rm$. The result is  reported in   Fig. \ref{fig:graph_R_min_vs_s_linear}. The critical coupling is obtained at the minimum for  $s/L\simeq $ 0.11 for $kd\simeq 0.17$, corresponding to a
subwavelength ratio $\lambda/(d+e) \simeq 18$. 
The low reflection is further illustrated in the insets where we report the measured  patterns of $|\eta(x,y)|$ for  $|\Rm|\simeq 0.5$ at $s/L=$ 0.48 and for $|\Rm|\simeq 2.10^{-2}$ at $s/L=$ 0.11.

To interpret the findings of the Fig. \ref{fig:graph_R_min_vs_s_linear}, it is sufficient to come back to \eqref{eqR}.
Adopting a  representation of $R$ in the complex map of $k$, a zero of $R$ appears at some distance of the real axis. The plot reports a typical case for the highest $s/L$ value (high leakage); in this case, the radiative damping is higher than the viscous damping. Next, decreasing  $s/L$ makes the radiative damping to decrease up to balance the viscous damping (yellow point realizing the critical damping). Eventually decreasing further the radiative damping moves the zero of $|R|$ in the lower half plane. 

\begin{figure}[h!]
	\centering
	\includegraphics[width=.85\columnwidth]{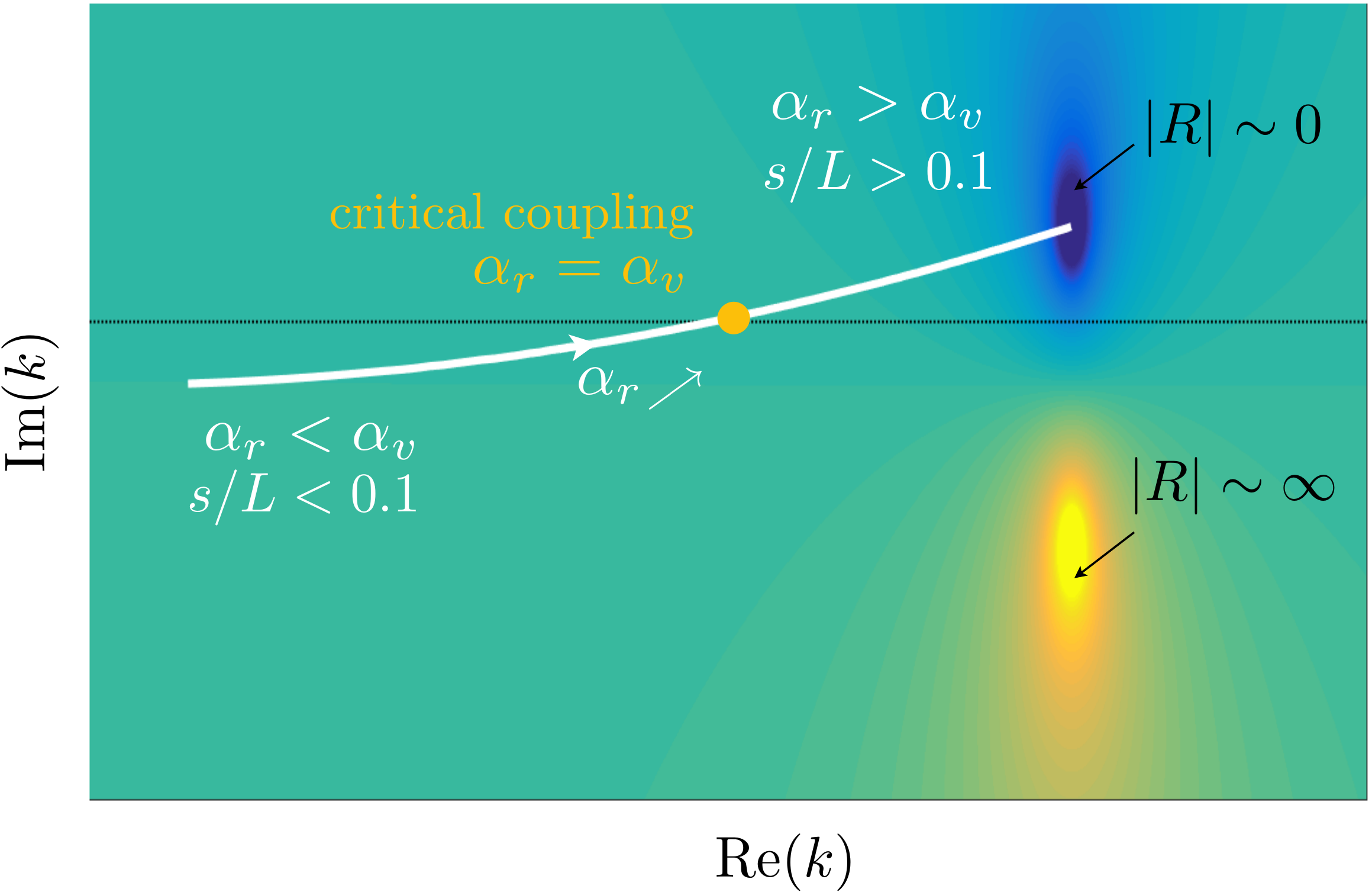}
	 \caption{Trajectories  of the zeros of reflection in the complex $k$-plane varying the radiation damping $\al$ increasing with $s$, for a fixed viscous damping (that of the linear regime); the complex resonance frequency is shifted according to results in Fig. \ref{figreso}. The critical coupling is achieved when the trajectory crosses the real axis (white point at $s/L\sim 0.1$.}
	\label{test}
\end{figure}

\vspace{.3cm}

\noindent{\em Critical coupling tuned by nonlinearities --}
It has been shown that water waves interacting with obstacles in the nonlinear regime experience an effective damping higher than that predicted in the linear regime, see {\em e.g.}   \cite{warnitchai_modelling_1998}. This is attributable to the energy taken by  structures generated in the fluid, and the stronger are the nonlinearities the higher is the effective damping within \cite{warnitchai_modelling_1998} a linear dependence  \cite{morison1953experimental}. 
Hence, it is possible to increase the internal damping and to balance it  with the radiative damping by increasing  the amplitude of the waves. 
In other words, for the family of resonators which are strongly coupled to the exterior, it is possible  to realize perfect absorption tuned by the nonlinearities.  
This critical coupling by nonlinear losses has been demonstrated recently for acoustic waves \cite{achilleos2016coherent}.

\begin{figure}[h]
			\centering
	\includegraphics[width=1\columnwidth]{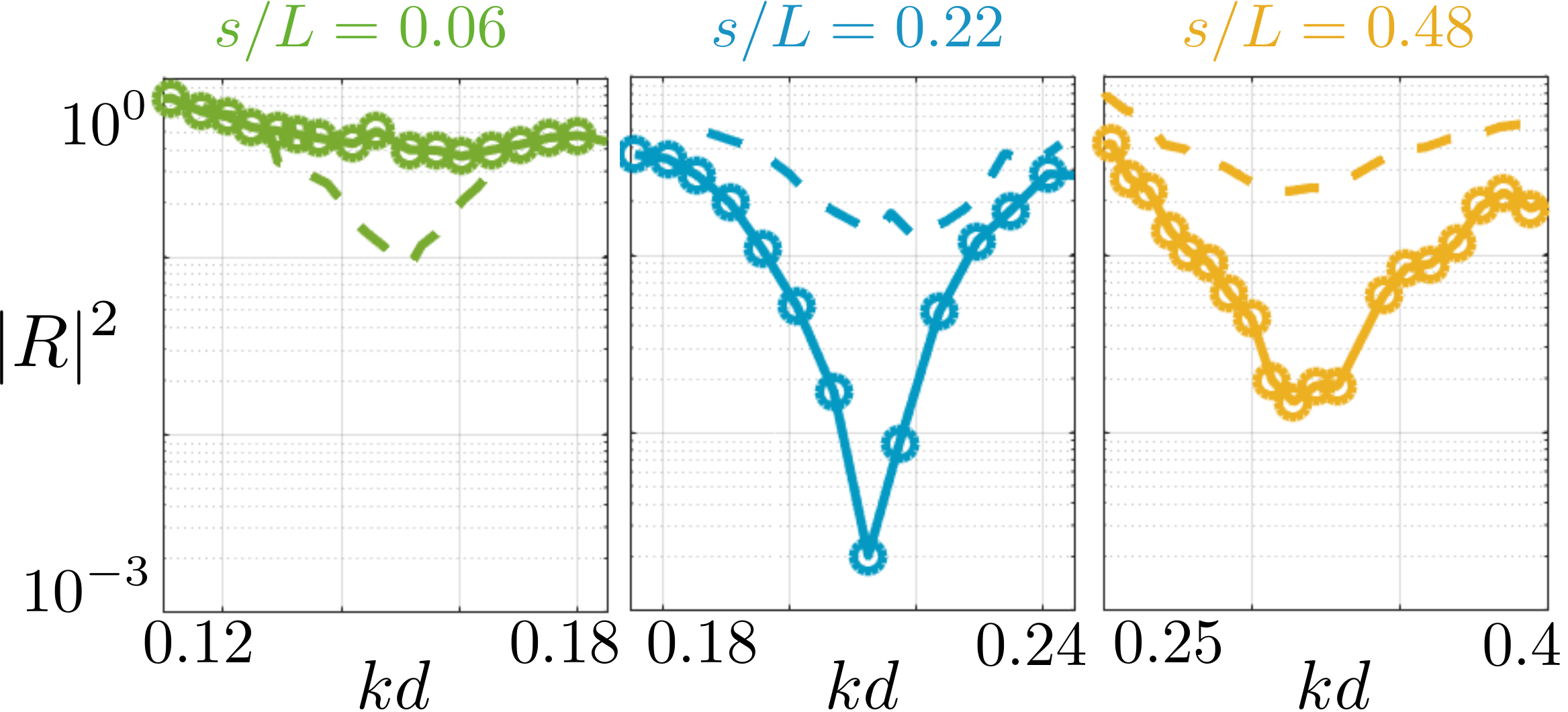}
	 \caption{Reflection coefficient $|R|^2$ as a function of the nondimensional wavenumber $kd$ for different entrance width $s/L$ and incident wave amplitude. Dashed line: linear regime; Circles: nonlinear regime.}
	\label{fig:graph_reflection_linear_nonlinear_Log}
\end{figure}

We consider resonators with relative openings $s/L=0.06$, $s/L=0.22$ and $s/L=0.48$.  We measure the reflectivity curves in the linear regime ($a=0.5$ mm) and in the non linear regime (with $a=2$ mm for $s/L=0.22$ and  $a=4$ mm for $s/L=0.06$ and 0.48); results are reported in  Fig. \ref{fig:graph_reflection_linear_nonlinear_Log}. 
For  $s/L=0.06$,   the reflectivity increases from the linear to  the non linear regimes. 
This is expected since we already know from the study in the linear regime that for such 
weakly coupled resonator, the radiative  damping was already weaker than the viscous damping in the linear regime
(see Fig. \ref{test}); 
 hence increasing this latter makes the things worse. Reversely,  for  $s/L=0.22$ and $0.48$, the reflectivity reduces a fact also expected from the Fig. \ref{test}. 

\begin{figure}[h!]
		\centering
	\includegraphics[width=.9\columnwidth]{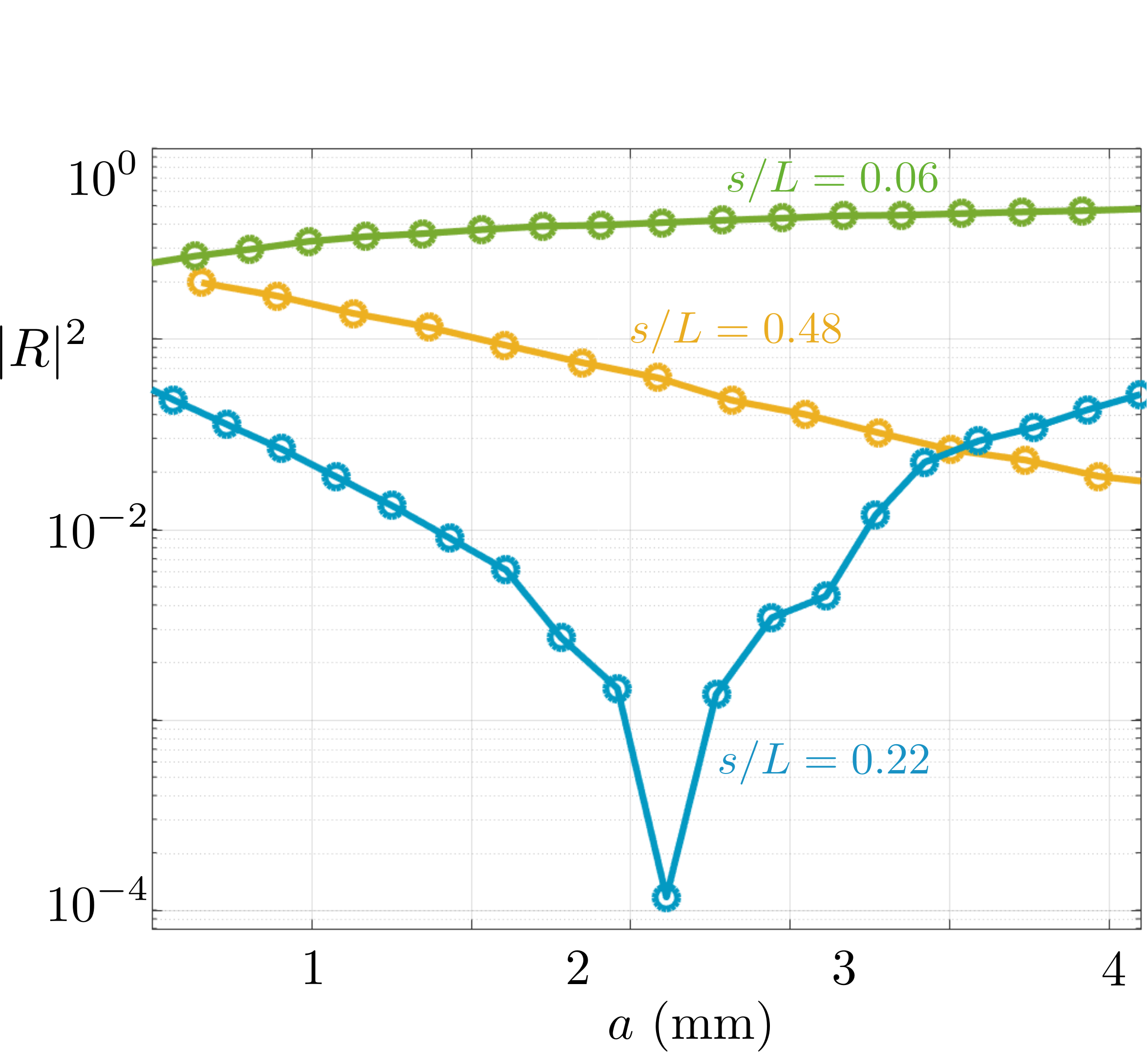}
	 \caption{Reflection coefficient as a function of the incident wave amplitude for different entrance widths $s/L$.}
	\label{fig:graph_reflection_vary_amp_aDd}
\end{figure}

\begin{figure}[h!]
		\centering
	\includegraphics[width=1\columnwidth]{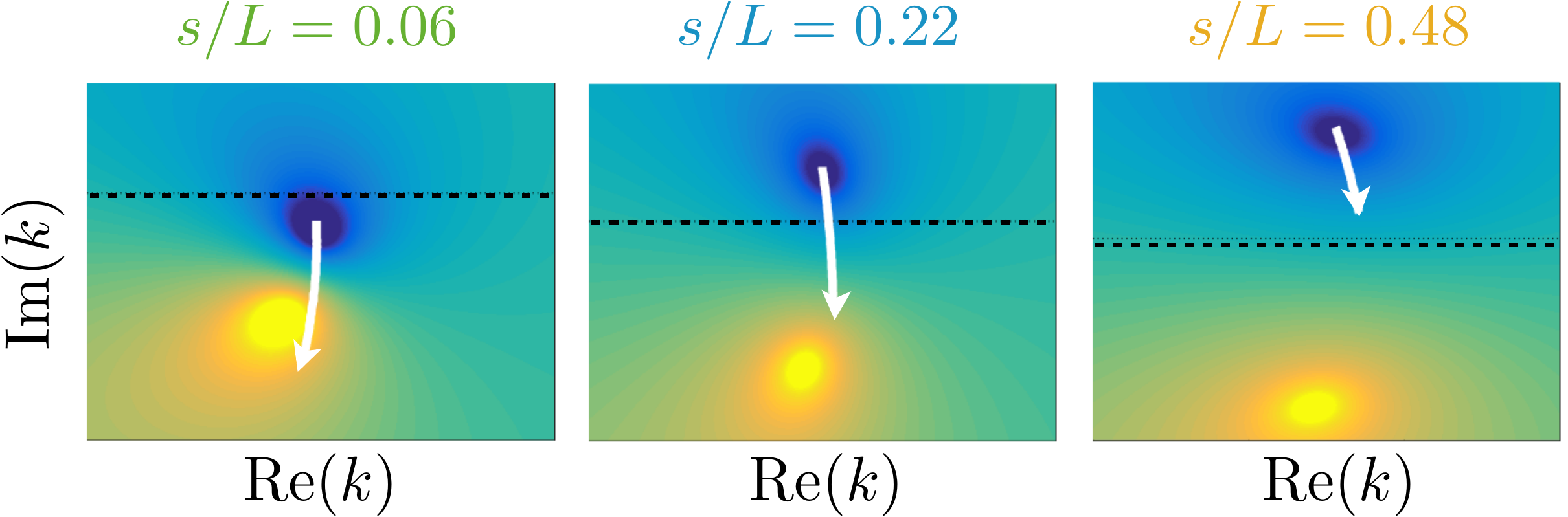}
	 \caption{Typical trajectories of the zero of reflectivity in the complex case, corresponding to our 3 experimental cases. }
	\label{fig:model_absorption_1D_vary_amp}
\end{figure}

To go further, we modify with constant step the wave amplitude in the nonlinear regime up to $a=4.5$ mm (the wavemaker is controlled precisely with an amplitude step of 0.1 mm). We report $|R|^2$ as a function of the wave amplitude $a$ in  Fig. \ref{fig:graph_reflection_vary_amp_aDd}.  
This representation complements that of the Fig. \ref{fig:graph_reflection_linear_nonlinear_Log} at $\Rm$ ($kd=0.14$, 0.21 and 0.3, respectively). Expectedly, for the resonator weakly coupled with the exterior ($s/L=0.06$), the reflectivity increases with the nonlinearities since the system is getting away from the critical coupling. Next,  both resonators $s/L=0.22$ and 0.48 are sufficiently coupled to the exterior, hence  increasing the non linearities makes these systems to approach  the critical coupling. 
However, the critical coupling is reached only for the resonator at $s/L=0.22$. For $s/L=0.48$,   we see that we should increase further the nonlinearities to reach the critical coupling for that large opening. 
It is  worth noting that for increasing the leakage with large $s/L$ produces  low quality factor  resonances, a fact which is recovered in the reflectivity curve in Fig. \ref{fig:graph_reflection_linear_nonlinear_Log} with a decrease in $|R|^2$ occurring in a broad frequency range. 
Eventually, the scenario described above is illustrated further in Fig. \ref{fig:model_absorption_1D_vary_amp} where the trajectory of the zeros of $|R|$ in the complex $k$-plane is illustrated in the 3 cases. 

\vspace{0.2cm}
In this work, we have demonstrated experimentally that we can obtain perfect wave absorption by tuning the geometry of a resonator, and consequently the radiation damping generated by this device. 
In addition, we demonstrated that the perfect absorption can be reached by tuning the incident wave amplitude (non linearity), which varies proportionally the intrinsic damping of the system. 
In this situation, when there is an excess of radiation damping (wide entrance to the resonator), by increasing the incident wave amplitude, we obtain broadband absorption due to the smaller quality factor of the resonance in this geometry.
A natural continuation of this work would be to apply it to an energy conversion system. 
In this case, in order to maximize the energy conversion, we have to minimize the intrinsic losses of the system (water viscous losses), keeping most of the mechanical resistance coming from the conversion device. 

%\section*{Acknowledgments}
\vspace{0.5cm}
E.M. acknowledges the support of CONICYT Becas Chile Doctorado. E.M., P.P., A.M. and V.P acknowledge the support Agence Nationale de la Recherche through the grant DYNAMONDE ANR-12-BS09-0027-01.

%%%%%%%%%%%%%%%%%%%%%%%%%%%%%%%%%%%%%%%%%%%%%%%%%%%%%%%%%
%%%%\bibliography{biblio1}

%

\end{document}